\documentclass[5p,times,twocolumn]{elsarticle}

\usepackage{amssymb}
\usepackage{lipsum}
\usepackage{amsmath,amsfonts}
\usepackage{algorithmic}
\usepackage{algorithm}
\usepackage{array}
\usepackage{textcomp}
\usepackage{stfloats}
\usepackage{url}
\usepackage{verbatim}
\usepackage{float}

\usepackage{color}

\begin{document}

\begin{frontmatter}

\title{Key Motifs Searching in Complex Dynamical Systems}

\author[first,second,third]{Qitong Hu}
\author[first,second,third]{Xiao-Dong Zhang\corref {cor1}}
\cortext[cor1]{Corresponding author: xiaodong@sjtu.edu.cn}
\affiliation[first]{organization={School of Mathematical Sciences, Shanghai Jiao Tong University},
            city={Shanghai},
            postcode={200240}, 
            state={Shanghai},
            country={China}}
\affiliation[second]{organization={Ministry of Education (MOE) Funded Key Lab of Scientific and Engineering Computing, Shanghai Jiao Tong University},
            city={Shanghai},
            postcode={200240}, 
            state={Shanghai},
            country={China}}
\affiliation[third]{organization={Shanghai Center for Applied Mathematics (SJTU Center), Shanghai Jiao Tong University},
            city={Shanghai},
            postcode={200240}, 
            state={Shanghai},
            country={China}}

\begin{abstract}
Key network motifs searching in complex networks is one of the crucial aspects of network analysis. There has been a series of insightful findings and valuable applications for various scenarios through the analysis of network structures. However, in dynamic systems, slight changes in the choice of dynamic equations and parameters can alter the significance of motifs. The known methods are insufficient to address this issue effectively. In this paper, we introduce a concept of perturbation energy based on the system's Jacobian matrix, and define motif centrality for dynamic systems by seamlessly integrating network topology with dynamic equations. Through simulations, we observe that the key motifs obtained by the proposed energy method present better effective and accurate than them by integrating network topology methods, without significantly increasing algorithm complexity. The finding of key motifs can be used to apply for system control, such as formulating containment policies for the spread of epidemics and protecting fragile ecosystems. Additionally, it makes substantial contribution to a deeper understanding of concepts in physics, such as signal propagation and system's stability.
\end{abstract}

\begin{graphicalabstract}
\includegraphics[scale=0.1]{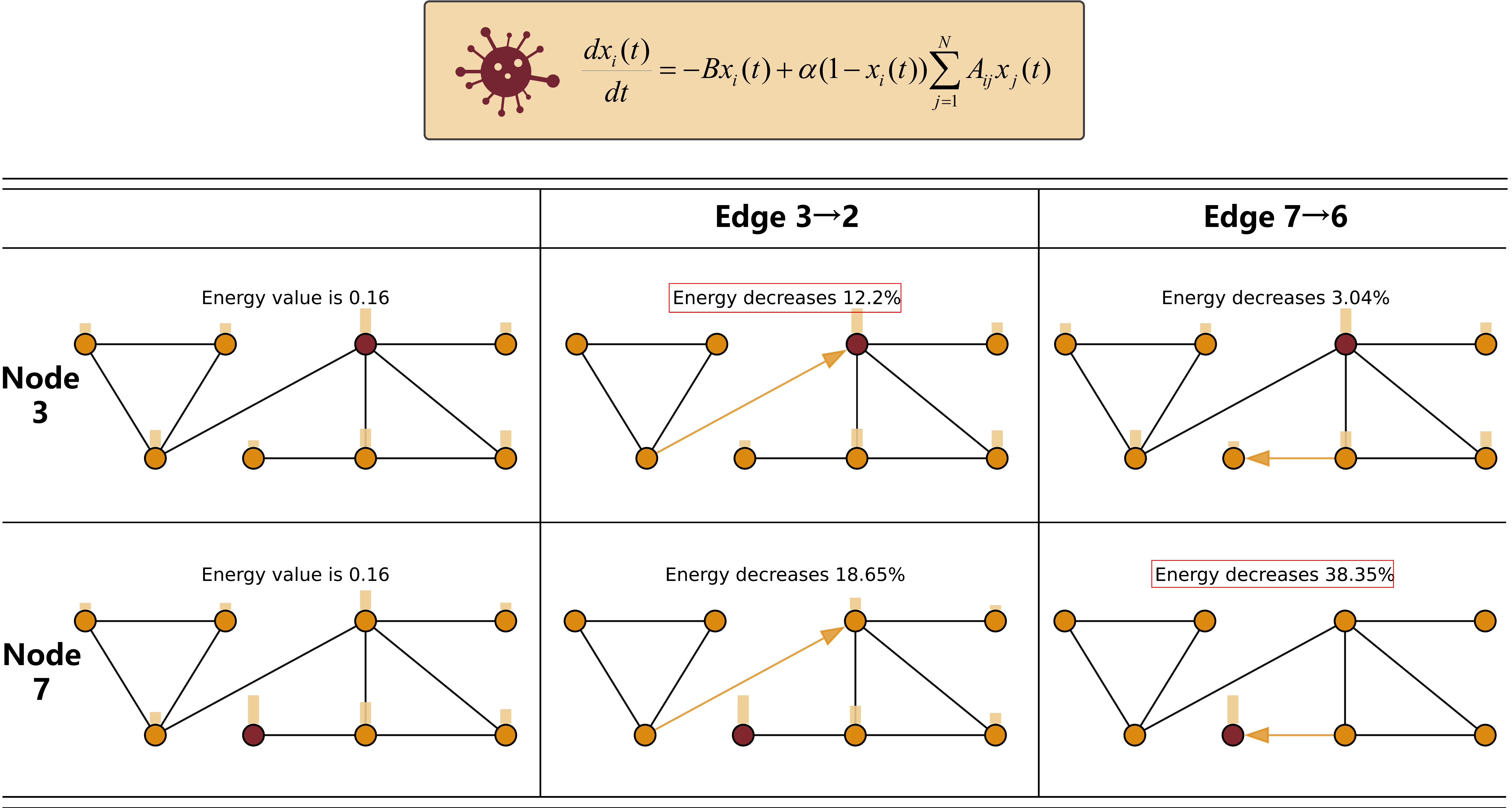}
\end{graphicalabstract}

\begin{highlights}
\item A new concept for perturbation energy is proposed, which is applicable to all types of perturbations.
\item Based on dynamical system and network structure, a new motif centrality and motif ranking algorithm is proposed,  which performs better effective and accurate than some known algorithms based on network topology. 
\item The proposed ranking algorithm can be applied for the control of spreading diseases, and offer a more flexible, scientific, and minimally disruptive containment policy compared to common lockdown policies.
\item The proposed ranking algorithm can be applied to the protection of ecosystems and the control of gene regulation, offering a more effective method to achieve specific goals.
\end{highlights}

\end{frontmatter}

\section{Introduction}
\par Dynamical systems serve as powerful tools for monitoring the evolution of real world systems over time, and extensive researches on dynamical systems has significantly enhanced our understanding of various physical phenomenon. For instance, the study of signal propagation contributes to our insights of dynamic processes like the spread of infectious diseases or violence\cite{Barzel2013UniversalityIN,Harush2017DynamicPO,Hens2019SpatiotemporalSP,Bao2022ImpactOB,HuPatterns2023,Bontorin2023ComplexBN,Thibeault2022TheLH}, while investigations into network stability and resilience deepen our awareness of ecological system tipping points\cite{Gao2016UniversalRP,Ma2021UniversalityON,Zhang2022EstimatingCD,Meena2020EmergentSI,ZHAO2024134126}. By leveraging specific dynamics properties, we can manipulate either the network structure or certain parameters within the network dynamics to achieve a specific objectives, referred to as network control\cite{Liu2011ControllabilityOC,Yan2015SpectrumOC,Sanhedrai2020RevivingAF,Sanhedrai2022SustainingAN,DSouza2023ControllingCN,YANG2022133499}, such as restoring system stability or reaching a particular fixed state. In this paper, we concentrate on control of the network structure and explore how motifs affect the properties of dynamical systems.

\par Network motifs, which can be categorized into undirected motifs and directed motifs, are fundamental building blocks of complex networks\cite{Milo2002NetworkMS}. Some common motifs, such as edges\cite{Girvan2001CommunitySI,Brhl2019CentralitybasedIO}, nodes\cite{Lu2016VitalNI,Liu2018OptimizingPC}, and cycles\cite{Bianconi2002NumberOL,Kim2005CyclicTI,Fan2020CharacterizingCS,Jiang2023SearchingFK}, all of which can significantly impact network properties and dynamics properties\cite{Lambiotte2019FromNT,Battiston2021ThePO,Battiston2020NetworksBP,StOnge2021UniversalNI}. A comprehensive exploration of network motifs is essential for understanding dynamics, particularly in identifying motifs with significant impacts on dynamic properties. This offers a new approach to comprehend network structure and dynamic behavior. Furthermore, such research holds significant importance in various fields, including dynamic system control\cite{Lizier2012InformationSL}, infectious disease prevention and control\cite{Fan2020CharacterizingCS}, information dissemination\cite{Bao2022ImpactOB,HuPatterns2023}.

\par From now on, numerous significant conclusions have been drawn regarding the centrality of network motifs. Among these, node centrality is the most extensively studied, encompassing aspects such as local topological properties, global topological properties, spectral properties, and propagation properties\cite{Lu2016VitalNI}. Various node centrality measures are employed depending on different cases and objectives to achieve specific goals. In recent studies, more interesting concepts and algorithms have emerged. For example, Changjun Fan et al. introduced reinforcement learning to identify key players in complex network\cite{Fan2020FindingKP}. Additionally, Siyang Jiang et al. used the Fiedler vector to rank cycles in network\cite{Jiang2023SearchingFK}, which is somewhat related to community detection. All of these findings have significant implications for controlling network dynamics.

\begin{figure*}[tbp]
\centering
\includegraphics[scale=0.1]{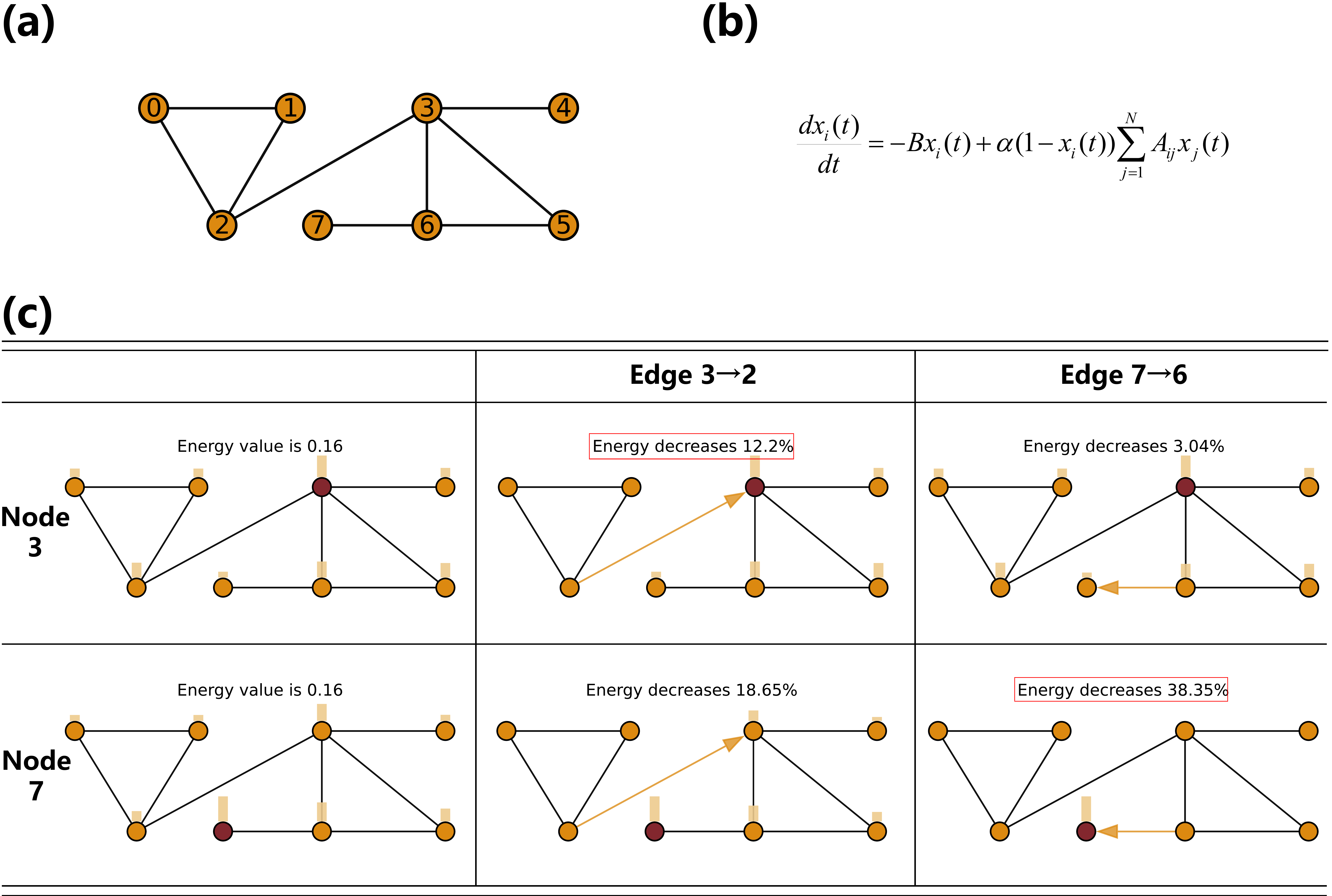}
\caption{\label{figure:1}\textbf{A static motif centrality is not available in dynamical system.} \textbf{(a)} A toy model with 8 nodes and 9 edges. \text{(b)} Epidemic dynamics form Eq.(\ref{equ:E:1}) with $B=0.8,\alpha=0.3$. \textbf{(c)} Simulation results are presented for the network proposed in (a) and the dynamics model proposed in (b). The first row depicts the outcomes on perturbing node $3$, while the second row shows the results for node $7$. \textbf{Noting that the direction of the arrow represents the direction of limited flow.} In the first column, the final state is displayed when inhibiting edge $3\to 2$, while the second column shows the effect of inhibiting edge $7\to 6$.}
\end{figure*}

\par {However, in dynamical systems, key nodes or edges are dynamically changing, as they are are influenced by the state of nodes and the interaction connections evolving them.} To illustrate this point, we consider a toy model and epidemic dynamics, and present the results in Figure \ref{figure:1}. In this simulation, we utilize the $2-$norm of the node state vector to represent the overall energy of the dynamical system, and apply fixed perturbations to different nodes, serving as perturbed nodes. It can be observed that the final state of the nodes varies depending on the selected perturbed node. Moreover, when restricting the same edge in two dynamical systems, we reach completely opposite conclusions: for the case with perturbed node $3$, constraining the edge $3\to2$ leads to a greater decrease in state compared to constraining $7\to6$. Conversely, in the case with perturbed node $7$, the situation is reversed, and the impact of edge $7\to6$ varies significantly with different source node choices.{(Note that $i\to j$ represents $i$ is the source node and $j$ is the target node, and we can set value for $A_{ji}$(or (j,i)) to control this arrow.)} This underscores the inadequacy of solely considering edge centrality based on network structure without incorporating dynamic behavior, and emphasizes the necessity to devise the motifs centrality that comprehensively consider both dynamic equations and network topology.

\par In this paper, we introduce a centrality for identifying key motifs based on the inverse of the Jacobian matrix, which takes into account both network topology and dynamics and can determine the importance of motifs at a specific state.  Unlike fixed network centrality, our proposed centrality may vary depending on different dynamical models or selections of dynamical parameters. This allows us to dynamically and accurately characterize the system's properties. Furthermore, the Jacobian matrix is a fundamental concept in dynamical systems and plays a crucial role in the studying network stability and signal propagation. It captures the linear variation properties that the node states satisfy at a given time. Our in-depth examination of the Jacobian matrix enhances our understanding of this mathematical concept, enabling us to further investigate the physical mechanisms of propagation and apply them to other physical domains.

\section{Framework on Motifs Searching for Dynamical Systems}
\subsection{Energy of Perturbation}
\par In this paper, we consider a dynamical system with $N$ nodes:
\begin{equation}
    \begin{aligned}
        \label{equ:dynamic:1}
        \frac{dx_i(t)}{dt}=F_i(\mathbf{x}(t),A), (i=1,\cdots,N).
    \end{aligned}
\end{equation}
in which $[F_i]_{i=1}^N$ represents a set of nonlinear functions, and $A$ denotes a general interaction matrix, encompassing both binary matrices and weighted matrices. For simplicity,
we can divide this system into self-dynamics and dynamical interaction, expressed as {$F_i(\mathbf{x}(t),A)=F(x_i(t))+\sum\limits_{j=1}^NA_{ij}G(x_i(t), x_j(t))$}\cite{Gao2016UniversalRP}. Moreover, the interaction between node $i$ and its neighborhood nodes can be simplified as $G(x_i(t), x_j(t))=H_1(x_i(t))H_2(x_j(t))$\cite{Hens2019SpatiotemporalSP}. The general dynamical model proposed in Eq.(\ref{equ:dynamic:1}) can unveil various dynamics model based on different choices of nonlinear functions, such as epidemic dynamics, biochemical dynamics, etc.
\par The stationary state in the unperturbed dynamical system, denoted as $[x_i^*]_{i=1}^N$, is obtained by setting the left hand of Eq.(\ref{equ:dynamic:1}) to be $0$, i.e. $F(\mathbf{x}^*,A)=0$. The signal propagation process is defined by introducing a perturbation $\Delta x_m(t)$ to the stationary state of a node $m$, and the shifted state of node $m$ is characterized by $x_m(t,m)=x_m^*+\Delta x_m(t)$. Furthermore, we can separate its initial state and dynamics equation by letting $\frac{d\Delta x_m(t)}{dt}=f(t)$ and $\Delta x_m(0)=\Delta x_m$. (It's important to note that the perturbation $\Delta x_m(t)$ in this article is not independent from time $t$, and the signal propagation proposed in the previous research\cite{Hens2019SpatiotemporalSP,Bao2022ImpactOB,HuPatterns2023} can be achieved by setting $\Delta x_m(t)=\Delta x_m$). This perturbation can drive nodes in this systems to their own shifted states $\mathbf{x}(t,m)=\mathbf{x}^*+\Delta \mathbf{x}(t,m)$. Furthermore, the perturbation for shifted states $\mathbf{x}(t,m)$ satisfies the following dynamic equation, as obtained from \cite{Bao2022ImpactOB,HuPatterns2023,Meena2020EmergentSI}:
\begin{equation}
    \begin{aligned}
        \label{equ:dynamic:2}
        \left\{
        \begin{array}{l}
        \frac{d\Delta \mathbf{x}(t,m)}{dt}=\tilde{\mathbf{J}}\Delta \mathbf{x}(t,m)+\mathbf{f}(t),\\
        \Delta \mathbf{x}(0,m)=\Delta x_m(0)e_m
        \end{array}
        \right.
    \end{aligned}
\end{equation}
in which $\mathbf{f}(t)=f(t)e_m$ represents the controlling vector for node $m$, {where $e_m$ denotes the $m-$th column of the identity matrix}. $\tilde{\mathbf{J}}$ is the perturbed Jacobian matrix, obtained by replacing the entries in the $m$-th rows of $\mathbf{J}$ with zero\cite{HuPatterns2023}. The Jacobian matrix $\mathbf{J}$ is defined as $\mathbf{J}=\left[\frac{\partial F_i(\mathbf{x}^*)}{\partial x_j}\right]_{i,j=1}^N$, and the perturbed Jacobian matrix $\tilde{\mathbf{J}}$ can be expressed as $\tilde{\mathbf{J}}=\left(I-e_me_m^T\right)\mathbf{J}$.
\par According to the Laplace transformation method outlined in Hu et al.'s work\cite{HuPatterns2023}, the solution for Eq.(\ref{equ:dynamic:1}) can be expressed using a matrix exponential function $\Delta \mathbf{x}(t)=e^{\tilde{\mathbf{J}}t}\Delta \mathbf{x}(0)$. $\Delta \mathbf{x}(t,m)$ will converge to a non-zero
constant vector $\Delta \mathbf{x}(\infty,m)=[\Delta x_i(\infty,m)]_{i=1}^N$\cite{HuPatterns2023}, referred to as the shifted state of node $i$. {This constant can be obtained using the limiting theorem of the Laplace transform for Eq.(\ref{equ:dynamic:2})}, and there is
\begin{equation}
    \begin{aligned}
        \label{equ:dynamic:3}
        \Delta \mathbf{x}(\infty,m)=\lim_{s\to 0}s\left(sI-\mathbf{J}+e_me_m^T\mathbf{J}\right)^{-1}\left[\Delta \mathbf{x}(0)+G(s)\right]e_m.
    \end{aligned}
\end{equation}
where $s$ is the corresponding variable to time $t$ and $\mathcal{L}[\mathbf{f}(t)]=\mathcal{L}[f(t)]e_m=G(s)e_m$. Utilizing the Sherman-Morrison formula $(A+uv^T)^{-1}=A^{-1}-\frac{A^{-1}uv^TA^{-1}}{1+v^TA^{-1}u}$ and considering the fact $\lim\limits_{s\to 0}G(s)=\int_0^{\infty}f(t)dt$, Eq.(\ref{equ:dynamic:3}) can be simplified as 
\begin{equation}
    \begin{aligned}
        \label{equ:dynamic:3.5}
        \Delta \mathbf{x}(\infty,m)
        &=\frac{\mathbf{J}^{-1}e_m}{(\mathbf{J}^{-1})_{mm}}\Delta x_m(\infty),
    \end{aligned}
\end{equation}
in which, according to the dynamical equation for $\Delta x_m(t)$, we have $\Delta x_m(\infty)=\Delta x_m(0)+\int_{0}^\infty f(t)dt$, and $A_{mm}$ represents the value of matrix $A$ on the $m$-th row and $m$-col, equivalent to $e_m^TAe_m$ {(the detailed proof has been provided in the Appendix A.1).} Then, We then define the energy $\mathcal{E}(m)$ of perturbation on node $m$ as the norm of $\Delta \mathbf{x}(\infty,m)$(introduced in Eq.(\ref{equ:dynamic:3.5})), i.e.
\begin{equation}
    \begin{aligned}
        \label{equ:dynamic:4}
        \mathcal{E}(m)=\Vert\Delta\mathbf{x}(\infty,m)\Vert_2^2&=\frac{(\mathbf{J}^{-T}\mathbf{J}^{-1})_{mm}}{(\mathbf{J}^{-1})_{mm}^2}\Delta x_m^2(\infty).
    \end{aligned}
\end{equation}
Through the definition of energy, if we assume the Jocabian matrix $\mathbf{J}$ is symmetric, then $\mathcal{E}(m)$ can be expressed as $\mathcal{E}(m)=\frac{(\mathbf{J}^{-2})_{mm}}{(\mathbf{J}^{-1})_{mm}^2}\Delta x_m^2(\infty)$. This specific case is analogous to expectation value and variance of random variables. The discrepancy between numerator and denominator is introduced by values in the $m$-th row and the $m$-th col of $\mathbf{J}$.
\subsection{Centrality of Motifs}
\par Through the definition of energy of perturbation proposed in Eq.(\ref{equ:dynamic:4}), each edge in the network contributes to this value(Note that here an edge here is considered as a directed arrow). The energy for the entire network arises from interactions among the energies associated with each edge. Therefore, it's not feasible to directly isolate the energy for only one edge from this intricate interaction. Here, we present an error analysis method, which involves introducing a small perturbation value $\varepsilon$ to a set of edges $E$. This perturbation affects the Jacobian matrix $\hat{\mathbf{J}}(E)=\mathbf{J}-\Delta\mathbf{J}(E)$, where $\Delta \mathbf{J}(E)=\varepsilon \sum\limits_{(i,j)\in E}\mathbf{J}_{ij}e_ie_j^T$, resulting in a perturbed energy denoted as $\mathcal{E}(m,E)=\Vert\Delta \hat{\mathbf{x}}(\infty,m,E)\Vert_2^2$. For simplicity in calculation, we consider two matrices: $\Delta G_1(E)=\mathbf{J}^{-T}\mathbf{J}^{-1}-\hat{\mathbf{J}}^{-T}(E)\hat{\mathbf{J}}^{-1}(E)$ and $\Delta G_2=\mathbf{J}^{-1}-\hat{\mathbf{J}}^{-1}(E)$, corresponding to the numerator and the denominator of Eq.(\ref{equ:dynamic:4}). Through some calculations, the change in energy $\Delta \mathcal{E}(m,E)$ between the original energy $\mathcal{E}(m)$ and the perturbed energy $\mathcal{E}(m,E)$ can be obtained as{(the detailed proof has been provided in the Appendix A.2)}:
\begin{equation}
    \begin{aligned}
        &\label{equ:energy:1}
        \Delta \mathcal{E}(m,E)=\mathcal{E}(m)-\mathcal{E}(m,E)
        \\
        =&\frac{(\mathbf{J}^{-T}\mathbf{J}^{-1})_{mm}}{(\mathbf{J}^{-1})_{mm}^2}\left(\frac{(\Delta G_1(E))_{mm}}{(\mathbf{J}^{-T}\mathbf{J}^{-1})_{mm}}-2\frac{(\Delta G_2(E))_{mm}}{(\mathbf{J}^{-1})_{mm}}\right)\Delta x_m^2(\infty),
    \end{aligned}
\end{equation}
where the relative difference can be calculated from Eq.(\ref{equ:energy:1}), and
\begin{equation}
    \begin{aligned}
        \label{equ:energy:2}
        \frac{\Delta \mathcal{E}(m,E)}{\mathcal{E}(m)}
        &=\frac{(\Delta G_1(E))_{mm}}{(\mathbf{J}^{-T}\mathbf{J}^{-1})_{mm}}-2\frac{(\Delta G_2(E))_{mm}}{(\mathbf{J}^{-1})_{mm}}.
    \end{aligned}
\end{equation}
\par Next, we estimate the values of $\Delta G_1(E)$ and $\Delta G_2(E)$ using the error analysis method. Firstly, according to the definition of the inverse matrix, we have $(\mathbf{J}-\Delta \mathbf{J})(\mathbf{J}^{-1}-\Delta G_2(E))=I$. Here, we assume $\mathbf{J}$ and $\mathbf{J}^{-1}$ are matrices of order $\varepsilon^0$, so $\Delta \mathbf{J}$ and $\Delta G_2(E)$ are matrices of order $\varepsilon^1$. Then, we only consider order $\varepsilon^1$ and obtain $\Delta G_2(E)=-\mathbf{J}^{-1}\Delta \mathbf{J}\mathbf{J}^{-1}$. Finally, by substituting the value of $\Delta\mathbf{J}(E)$ into $\Delta G_2(E)$, we can obtain the exact estimation for $(\Delta G_2(E))_{mm}$:
\begin{equation}
    \begin{aligned}
        \label{equ:energy:3}
        (\Delta G_2(E))_{mm}=-\varepsilon \sum\limits_{(i,j)\in E}\mathbf{J}_{ij}(\mathbf{J}^{-1})_{mi}(\mathbf{J}^{-1})_{jm}.
    \end{aligned}
\end{equation}
Similarly, according to the definition of inverse matrix, we have $(\mathbf{J}-\Delta \mathbf{J})(\mathbf{J}-\Delta \mathbf{J})^T(\mathbf{J}^{-T}\mathbf{J}^{-1}-\Delta G_1)=I$. Here, $\Delta \mathbf{J}$, $\Delta \mathbf{J}^T$, $\Delta G_1(E)$ in this formula are matrices of order $\varepsilon^1$. Then, we ignore terms of order $\varepsilon^2$ and higher, and obtain $\Delta G_1=-\mathbf{J}^{-T}\mathbf{J}^{-1}(\Delta \mathbf{J}\mathbf{J}^T+\mathbf{J}\Delta \mathbf{J}^T)\mathbf{J}^{-T}\mathbf{J}^{-1}$. Finally, by substituting the value of $\Delta\mathbf{J}(E)$ into $\Delta G_1(E)$, and we can obtain the exact estimation for $(\Delta G_1(E))_{mm}$:
\begin{equation}
    \begin{aligned}
        \label{equ:energy:4}
        \Delta G_1(E)_{mm}
        &=-2\varepsilon\sum\limits_{(i,j)\in E}\mathbf{J}_{ij}(\mathbf{J}^{-T}\mathbf{J}^{-1})_{mi}(\mathbf{J}^{-1})_{jm}.
    \end{aligned}
\end{equation}
In this paper, we provide a definition of centrality of motifs $\mathcal{S}(m,E)$ under the concept of energy of perturbation, and let {$\mathcal{S}(m,E)=\frac{\Delta \mathcal{E}(m,E)}{\varepsilon\cdot\mathcal{E}(m)}$}. Substituting the value of $\Delta G_1(E)_{mm}$ from Eq.(\ref{equ:energy:4}) and value of $\Delta G_2(E)_{mm}$ from Eq.(\ref{equ:energy:3}) into Eq.(\ref{equ:energy:2}), there should be
\begin{equation}
    \begin{aligned}
        \label{equ:energy:5}
        \mathcal{S}(m,E)
        =2\sum\limits_{(i,j)\in E}\mathbf{J}_{ij}(\mathbf{J}^{-1})_{jm}\left(\frac{(\mathbf{J}^{-1})_{mi}}{(\mathbf{J}^{-1})_{mm}}-\frac{(\mathbf{J}^{-T}\mathbf{J}^{-1})_{mi}}{(\mathbf{J}^{-T}\mathbf{J}^{-1})_{mm}}\right).
    \end{aligned}
\end{equation}
Therefore, we obtain the motif score $\mathcal{S}(m,E)$, representing the centrality of motif $E$.
\par Specifically for centrality of nodes, we can fix a node $i$ and select $N_j=\{i\to j,j\to i,A_{ij}=1\}$, then we can obtain
\begin{equation*}
    \begin{aligned}
        \mathcal{S}(N_i)
        =2\left(\frac{(\mathbf{J}^{-1}+\mathbf{J}^{-T})_{im}}{(\mathbf{J}^{-1})_{mm}}\delta_{im}-\frac{(\mathbf{J}^{-1})_{im}^2+(\mathbf{J}^{-T}\mathbf{J}^{-1})_{mi}}{(\mathbf{J}^{-T}\mathbf{J}^{-1})_{mm}}\right).
    \end{aligned}
\end{equation*}
\begin{algorithm}[H]
\caption{Motif Ranking in Dynamical System}\label{alg:alg1}
\begin{algorithmic}
\STATE 
\STATE {\textbf{Input}: Dynamical function $[F_i]_{i=1}^N$ , complex network $A$, motifs $E$ and time $t$.}
\STATE {\textbf{Output}: Ranking of motifs isomorphic to $\mathbf{E}$ at time $t$.}
\STATE \hspace{0.5cm} \textbf{1.} Calculate state $\mathbf{x}(t)$ at time $t$ through Eq.(\ref{equ:dynamic:1}).
\STATE \hspace{0.5cm} \textbf{2.} Calculate Jacobian matrix $\mathbf{J}$ at time $t$ through $\mathbf{x}(t)$.
\STATE \hspace{0.5cm} \textbf{3.} Select a motif $E$ isomorphic to $\mathbf{E}$.
\STATE \hspace{0.5cm} \textbf{4.} Calculate centrality $\mathcal{S}(E)$.
\STATE \hspace{0.5cm} \textbf{5.}  Back to 3. Until all the motifs have been ranked.
\STATE \textbf{The larger the centrality $\mathcal{S}(E)$, the more key the motifs.}
\end{algorithmic}
\label{alg1}
\end{algorithm}
\noindent If $i=m$, then $\mathcal{S}(N_m)=2\left(1-\frac{1}{\mathcal{E}(m)}\right)$, and the change in energy $\Delta \mathcal{E}(m,N_m)=2(\mathcal{E}(m)-1)$, this suggests that our energy of perturbation $\mathcal{E}(E)$ is also applicable for node centrality.
\par Through the estimation for $\mathcal{S}(E)$ proposed in Eq.(\ref{equ:energy:5}), we can determine the importance of motifs solely based on the Jacobian matrix. This approximation encapsulates information about perturbations, which can provide better predictions for dynamical system compared to static centrality, such as node centrality or eigenvector vector. Using our motif centrality, we can estimate key nodes, edges, or cycles in a dynamical system, the detailed algorithm has been provided in Algorithm \ref{alg:alg1}.
\begin{figure}[htbp]
\centering
\includegraphics[scale=0.1]{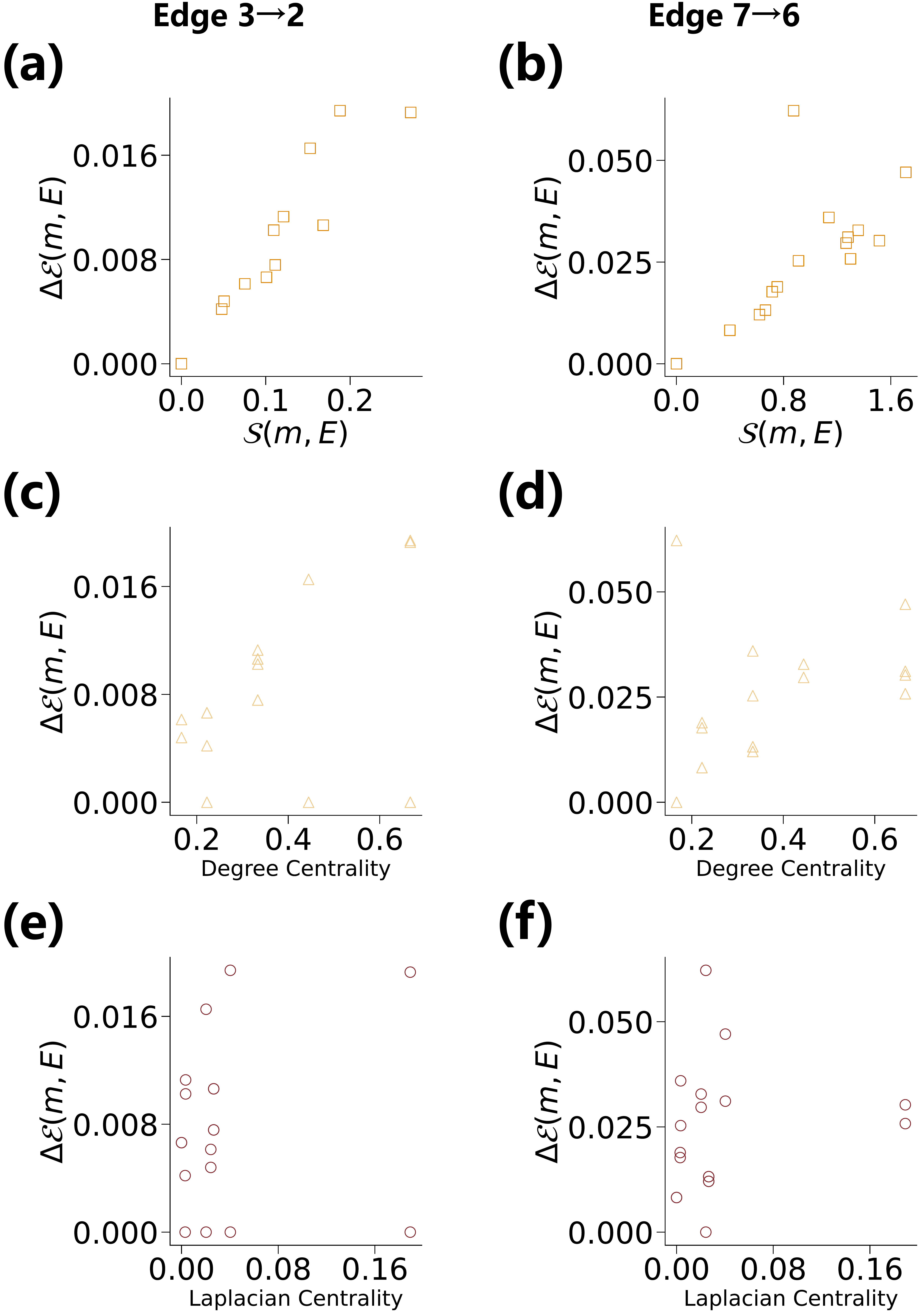}
\caption{\label{figure:2}\textbf{Comparison of different centrality and change in energy $\Delta \mathcal{E}(m,E)$ for different two choice of perturbed nodes}.\textbf{(a)} Comparison of our motif centrality $\mathcal{S}(m,E)$ and change in energy on network proposed in Figure \ref{figure:1} and epidemic dynamics proposed in Eq. (\ref{equ:E:1}) with $B=0.8,\alpha=0.01$.
\textbf{(b)} Degree centrality. \textbf{(c)} Laplacian centrality.}
\end{figure}
\par To highlight the advantages of our motif centrality, we compare two static centrality in complex network: the first is the degree centrality $\frac{d_id_j}{n\langle d\rangle}$, which calculates the expectation number of edges between node $i$ and $j$ based on homogeneous assumption. Here, $d_i=\sum\limits_{j=1}^NA_{ij}$ represents the degree of node $i$. The second is the Laplacian centrality, introduced in Jiang et al.'s work\cite{Jiang2023SearchingFK}, is defined as $\sum\limits_{(i,j)\in E}(\xi_2(i)-\xi_2(j))^2$, where $\xi_2$ is the Filder vector of the network. We preform simulations on the network depicted in Figure \ref{figure:1}, respectively perturbing on node $3$ and limiting edge $3\to 2$, and similarly for node $7$ and edge $7\to 6$, and investigate relationship between the motif centrality and the change in energy $\Delta \mathcal{E}(m,E)$. The change in energy can represent the signal propagated through motifs, especially if we select $\varepsilon=1$, it can interpreted as the effect of deleting one motif. The results are shown in Figure \ref{figure:2}, indicating that our motif centrality can better monitor the change in energy $\Delta \mathcal{E}(m,E)$ regardless of which node is perturbed, while the other two centrality do not have a clear correlation.

\par Furthermore, our motif centrality requires the values of the row and column of $\mathbf{J}^{-1}$ inferred in motifs set $E$, with a computational complexity of $\mathcal{O}(N^2)$. In comparison, the degree centrality is related to the number of edges in the network, with a complexity of $\mathcal{O}(Nd)$, where $d$ is the average degree of the network. The Laplacian centrality is related to the eigenvector corresponding to the second smallest eigenvalue of the Laplacian matrix, with a complexity of $\mathcal{O}(N^2)$. It is evident that our motif centrality does not have a significant disadvantage in algorithmic complexity and consistently outperforms the other two centrality in terms of results.

\section{Applications to Epidemic Dynamics}
\par Infectious diseases are illnesses caused by pathogens that can be transmitted among humans, animals, and between humans and animals. The spread of an infectious disease can have significant impacts on production, livelihoods, and the health of thousands of people. With the rapid development of society productive forces and transportation systems, the rate of spread for infectious diseases is expected to be more severe compared to ancient times. In the 21st century, China alone has witnessed several large-scale infectious disease outbreaks, such as the SARS epidemic in 2004\cite{Diseases2003AnAR,Wenzel2003ListeningTS} and the COVID-19 epidemic in 2021\cite{Estrada2020COVID19AS,Chinazzi2020TheEO,Aleta2020ADA}. To ensure the safety of the general population and maintain the normal functioning of the national economic, the governments and relevant departments must need to monitor and predict the spread of infectious diseases. This requires the assistance of mathematical tools to formulate effective policies for controlling.

\par In mathematics, epidemic dynamics are usually used to characterize the population changes in different groups during the spread of an infectious disease. Common epidemic dynamics include the $SI$, $SIS$, $SIR$, $SEIR$, where $S$, $I$, $E$ and $R$ respectively represents the proportion of the susceptible\cite{YAGASAKI2023133820,LIU2021132903}, exposed, infected and recovered individuals. These groups interact each other based on certain physical laws, undergoing transitions with specific probabilities. For example, susceptible individuals ($S$) can become exposed ($E$) through a certain reproduction number $R_0$, exposed individuals($E$) go through a latent period ($T_E$) to become infected ($I$), and infected individuals go through a recovery period ($T_I$) to become recovered ($R$)\cite{Aleta2020ADA,WANG2022133183}. It's important to note that the recovered individuals($R$) include both those who have recovered or died, and they do not revert to being susceptible, reflecting the realities of infectious diseases where recovery leads to short-term immunity.
\par The model for these four groups can depict the process of the spread of an epidemic in human society and play a crucial role in multiple epidemic predictions and control efforts. It provides strong theoretical support and data feedback for formulating epidemic controlling policies, allocating rescue materials, managing medical resources, and deciding on lockdown policies. In this paper, we focus on the change in susceptible individuals ($S$), with its general formula being\cite{Dodds2005AGM}
\begin{figure*}[htbp]
\centering
\includegraphics[scale=0.1]{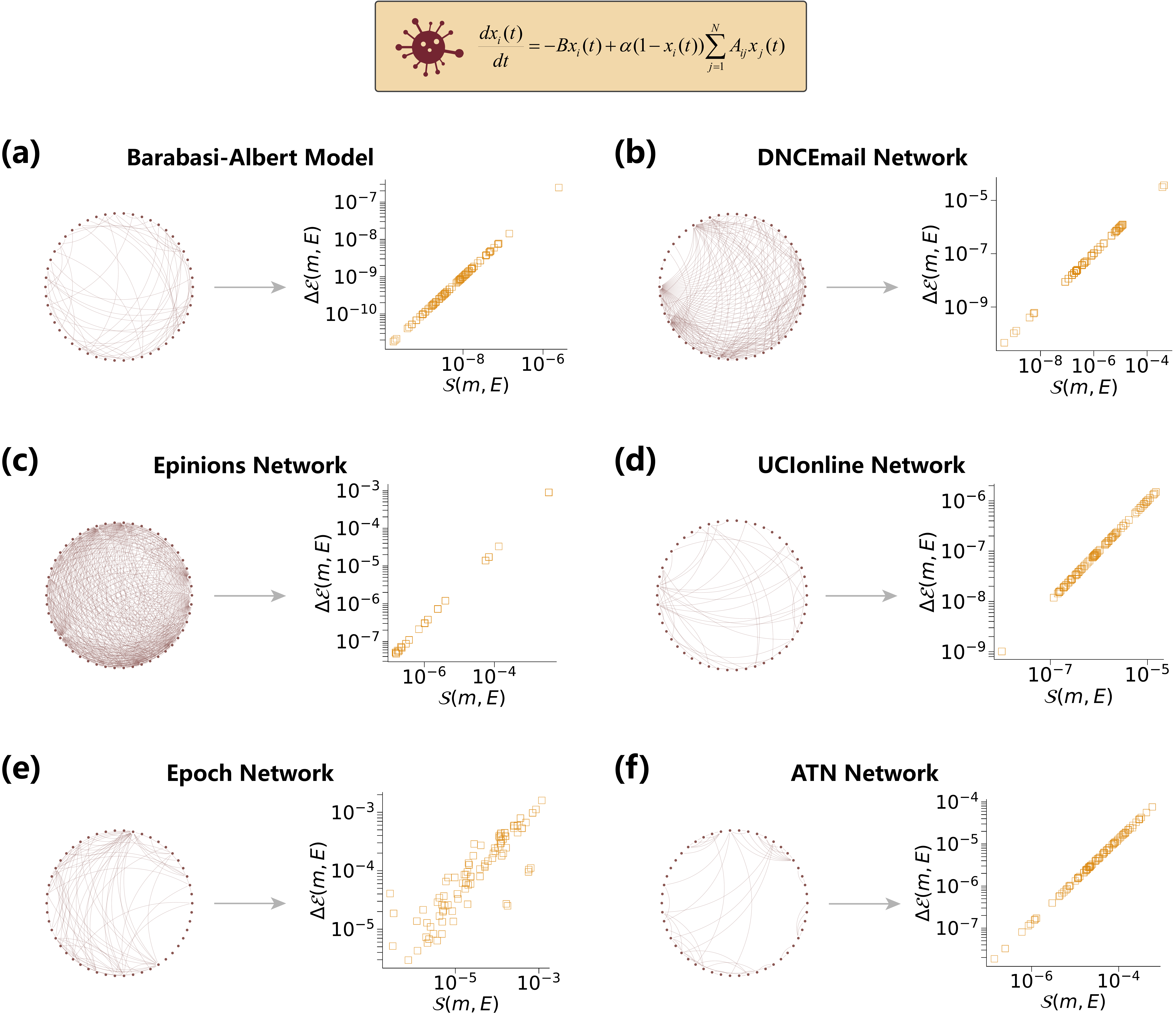}
\caption{\label{figure:3}\textbf{Our theoretical centrality $\mathcal{S}(m,E)$ v.s. change in Energy $\Delta \mathcal{E}(m,E)$ for fixed perturbed node on epidemic dynamics and different networks.}\textbf{(a)} Simulation on Barab\'asi-Albert model and epidemic dynamics proposed in Eq. (\ref{equ:E:1}) with $B=0.8,\alpha=0.01$. \textbf{(b)} DNCEmail network. \textbf{(c)} Epinions network. \textbf{(d)} UCIoline network. \text{(e)} Epoch network. \textbf{(f)} ATN network.}
\end{figure*}
\begin{equation}
    \begin{aligned}
        \label{equ:E:1}
        \frac{dx_i(t)}{dt}=-Bx_i(t)+\alpha(1-x_i(t))\sum_{j=1}^NA_{ij}x_j(t),
    \end{aligned}
\end{equation}
in which $x_i(t)$ is the ratio of susceptible
individuals at time $t$, $B$ is the recovery rate and $\alpha$ denotes the infection rate, corresponding respectively to $T_R$/$T_E$ and $R_0$. The actual value of $\alpha$ is influenced by the characteristics of disease. Infectious diseases transmitted through simple behaviors, such as airborne or respiratory transmission, often have a higher $\alpha$, as observed in cases like SARS and COVID-19. Conversely, diseases spread through complex behaviors like fluid or blood transmission tend to have a lower $\alpha$, as seen in diseases like HIV and syphilis.
\subsection{Verification of Theoretical Framework}
\par In the real world, the outbreak of infectious diseases often originates in one or a few cities and then spreads to other countries/area.  It is widely accepted that outbreak typically start from a single city. To limit the spread, governments need to implement relevant policies, including lockdowns and restrictions on travel between cities. In field of complex networks, lockdowns can be considered as removing all connections of a specific node with other nodes in the network, while restricting travel between cities can be seen as removing one or several edges in the network. Surprisingly, these two policies align well with the concept of network centrality and motif searching proposed in this paper, and the essential policy should be promulgated based on the motifs with the highest centrality.
\par To validate and refine our theoretical framework, we utilize epidemic dynamics to simulate the spread of diseases on real networks and conduct applied analyses in real-world cases. In this section, we employ changes in the energy of perturbation to monitor the spread and compare this value with our centrality proposed in Eq.(\ref{equ:energy:5}). Simulations are preformed on the Barab\'asi-Albert model and five other real networks: the first is DNCEmail network (DNCEmail), a network of emails from the 2016 democratic national committee email leak, consisting of 1834 nodes and 4367 edges\cite{DNCEmail2015}. The second is Epinions network (Epinions), a binary online social network consisting of 467 nodes and 6538 edges\cite{Tang2015TransferLP}. The third is UCIonline network (UCIonline), a collects real-time information networks of UC-Irvine students over a 218-day period, consisting of 1893 nodes and 27670 edges\cite{Opsahl2009ClusteringIW}. The fourth is Email epoch network (Epoch), a scale-free network collected email social networks over a 6-month period, consisting of 3185 nodes and 31885 edges \cite{Eckmann2004EntropyOD}. The fifth is Advogato trust network (ATN), a symmetric social network constructed from connections within the community of open source developers, consisting of 539 nodes and 23540 links \cite{Massa2009BowlingAA}. We removed the edges in the suggested epidemic networks and simulated the observed change in energy $\Delta \mathcal{E}(m,E)$. The simulation results, as depicted in Figure \ref{figure:3}, clearly demonstrate that our theoretical predictions align perfectly with the observed change in energy $\Delta \mathcal{E}(m,E)$ in real-world cases, this validates the correctness of our theoretical framework and the effectiveness of our motif centrality.
\subsection{Advice to Lockdown Policy}
\par In this section, we will present evidence for the effectiveness of our theoretical framework in disease control. After an outbreak of infectious disease, governments are limited by a certain timeframe to gather information about the epidemic and then implement corresponding control policy. Issuing reasonable policies for disease spread involves finding a balance between ensuring public safety and allowing continuous economic activities. For example, complete relaxation without any control policies poses a significant threat to public healthy, while overly strict control policies can hinder economic activities, severely impacting people's livelihoods and daily lives. Even a lockdown policy for a city can directly affects the lives of its residents and introduce instability factors into society. Hence, determining when and how to choose an appropriate level of control under scientific guidance and mathematical assistance become crucial. In this paper, we propose a framework for restricting interactions between two nodes/cities based on our proposed edge motif centrality, prioritizing pairs of nodes with the highest motif centrality. Compared to the rigid lockdown policies, our approach is more flexible, efficient, and ensures the normal functioning of daily life, which has significant practical value.
\par In our framework, we utilize the energy $\mathcal{E}(m,E)$ to gauge the severity of infectious disease spread, and mitigate it by selectively removing edges between network nodes. It is evident that the change in energy $\Delta \mathcal{E}(m,E)$ represents the importance of choice of edge set $E$. However, removing more edges will have a greater impact on the original societal activities. This prompts us to balance the relationship between the size of edge set $|E|$ and $\Delta \mathcal{E}(m,E)$, aiming to restrict disease spread while minimizing disruption to the 
\begin{figure}[tbp]
\centering
\includegraphics[scale=0.095]{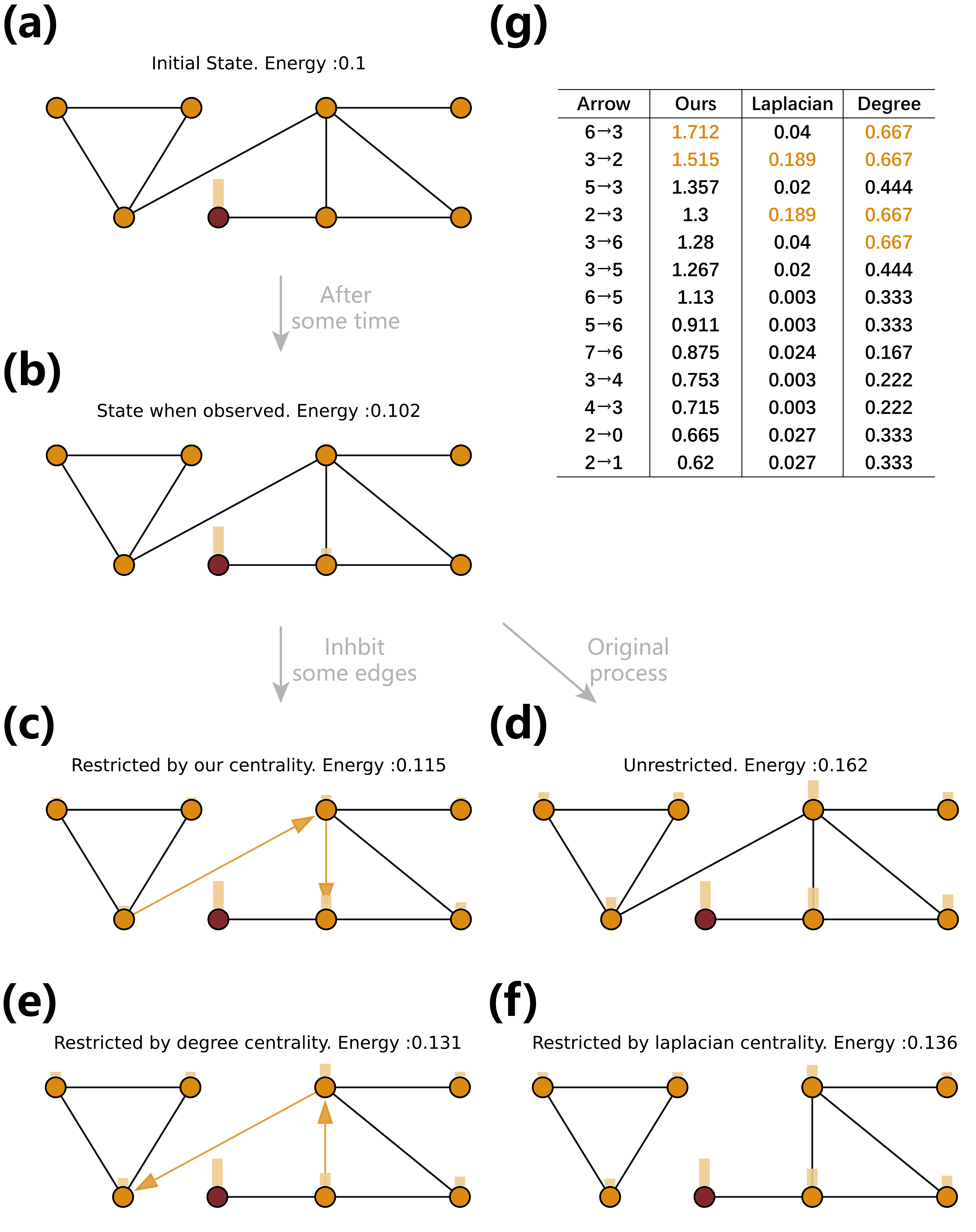}
\caption{\label{figure:4}\textbf{Inhibiting a small set of edge can deeply effectively decrease spreading level of infectious diseases.} Simulation on network proposed in Figure \ref{figure:1} and epidemic dynamics proposed in Eq.(\ref{equ:E:1}) with $B=0.8,\alpha=0.3$. \textbf{(a)} Initial state of dynamical system, with perturbed node $7$. \textbf{(b)} After a time period, the disease will spread to other nodes. \textbf{(c)} State if inhibiting two directed edges based on our centrality.\textbf{(d)} State if not inhibiting any edge. It can be observed that only inhibiting two directed edges can effectively decrease the spreading degree of diseases. \textbf{(e)} State if inhibiting two directed edges based on Laplacian centrality. \textbf{(f)} State if inhibiting two directed edges based on degree centrality. \textbf{(g)} Our centrality, Laplacian centrality and degree centrality are applied to select 13 directed edges, ranked by our centrality. The orange items represent the two highest values of one centrality.}
\end{figure}
overall network connectivity. We define efficiency for edge set $E$ as the ratio $\frac{\Delta \mathcal{E}(m,E)}{|E|}$, with the goal of identify the edge set $E$ with the highest efficiency. In mainstream views, if we could precisely identify the source, removing the connections of that node with the external network would be sufficient. However, due to the lag of information, deleting the connection to the perturbed node at this time may not be the most effective, as the disease has already spread throughout the network. This emphasizes the need for a robust guiding strategy. Through our motif ranking algorithm, as described in Algorithm \ref{alg:alg1}, we can assign centrality values to all edges using Eq.(\ref{equ:energy:5}), and obtain $E$ by selecting edges with the highest motif centrality $\mathcal{S}$. This algorithm allows us to select one edge at a time.

\begin{figure*}[tbp]
\centering
\includegraphics[scale=0.1]{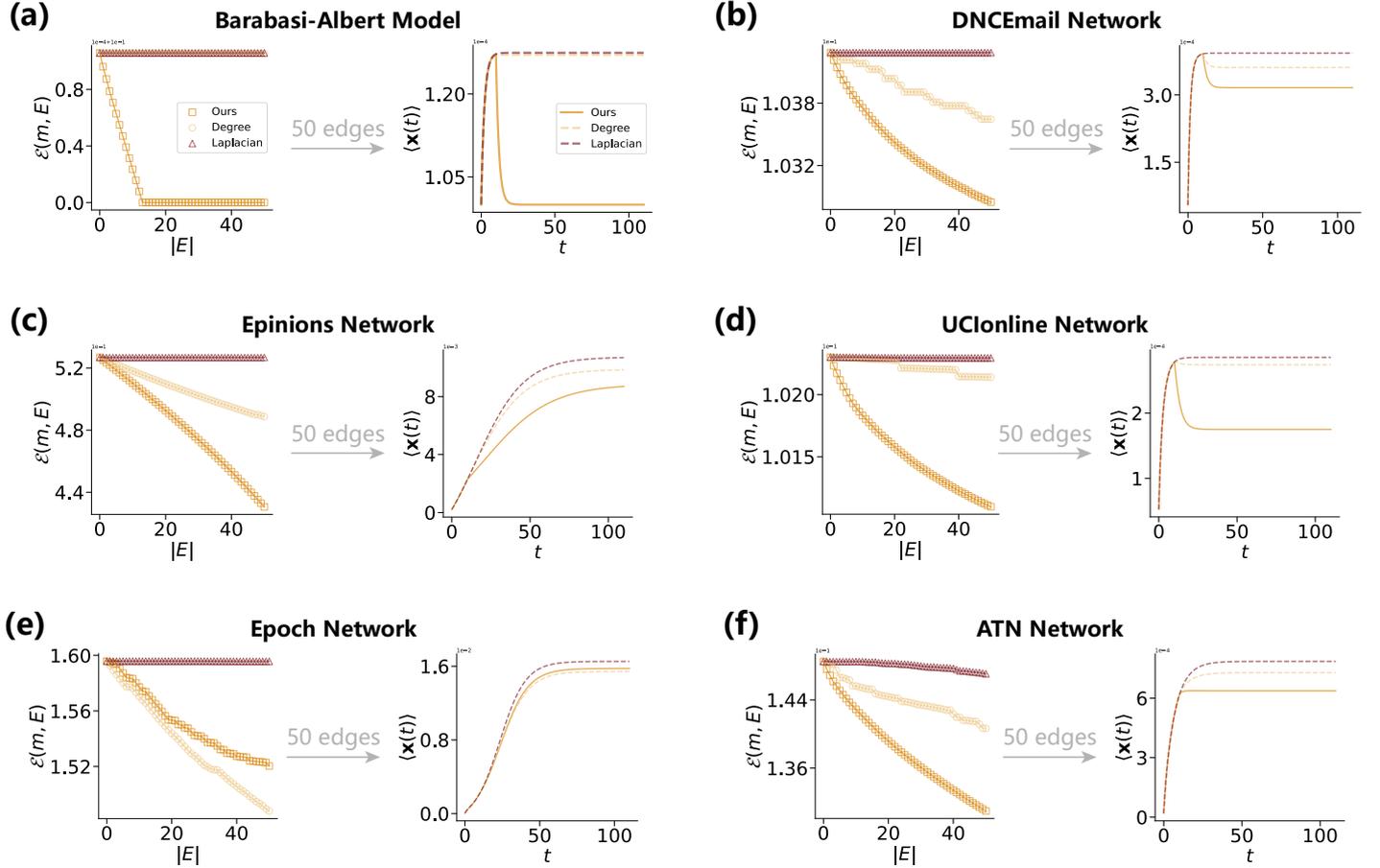}
\caption{\label{figure:5}\textbf{Effect of inhibiting edges ranked by different centrality on centrality of motifs $\mathcal{E}(m,E)$ and average state $\langle\mathbf{x}(t)\rangle$.} \textbf{(a)} Simulation on Barab\'asi-Albert model and epidemic dynamics proposed in Eq. (\ref{equ:E:1}) with $B=0.8,\alpha=0.01$. On the left-hand side, we observe the trend of motif energy $\mathcal{E}(m,E)$ over the size of $E$, denoted as $|E|$, under three different ranking algorithms: our centrality, degree centrality, and Laplacian centrality. On the right-hand side, we depict the trend of the average state $\langle\mathbf{x}(t)\rangle$ over time. Specifically, we consider the state when deleting 50 edges under different ranking algorithms. \textbf{(b)} DNCEmail network. \textbf{(c)} Epinions network. \textbf{(d)} UCIoline network. \text{(e)} Epoch network. \textbf{(f)} ATN network.}
\end{figure*}

\par To illustrate this concept, we preform an epidemic simulation on the network proposed in Figure \ref{figure:1}. After a certain period of continuous perturbation to a node in the network, we compare the final infection outcomes under different policies (Note that we examine the dynamic system after "a certain period of perturbation" to consider the time required for the government to gather specific information about the infectious disease and determine the policy). We present our results in Figure \ref{figure:4} by perturbing node $7$ and restricting edges {$6\to3$ and $3\to2$}. In comparison, we observe that the proportion of deleted directed edges in the entire network is only $11.1\%$, but the overall network energy decreases by $29.0\%$, and the network energy excluding the perturbed node decreases by $75.8\%$. The efficiency excluding the perturbed node is by $580\%$ times higher than deleting all edges! While For the two directed edges identified by the Laplacian centrality or degree centrality rankings, such as {$3\to2$ and $2\to3$} for Laplacian centrality, the final energy is much larger compared to the energy restricted by our centrality measure. This example indicates employing scientific control policies can enhance the reliability, rationality, and robustness of policy formulation, yielding better results compared to other centrality measures.
\par Furthermore, we conducted simulations on the theoretical network and five real networks mentioned in Figure \ref{figure:3}. We calculate the perturbed energy of system $\mathcal{E}(m,E)$ over size of $E$, and the average state of nodes $\langle \mathbf{x}(t)\rangle=\frac{1}{N}\sum\limits_{i=1}^Nx_i(t)$ after restricting certain edges over time $t$. All of these result are compared with both the degree centrality and Laplacian centrality, and are presented in Figure \ref{figure:5}. It is evident that our centrality can effectively decrease centrality of motif and can deeply affect average state of dynamical system. Moreover, our result consistently outperformed the other two centrality in almost all scenarios, and the significant reduction in control under a small number of restricted edges indicates that our centrality has a distinct advantage in addressing infectious disease prevention and control.
{
\begin{figure*}[tbp]
\centering
\includegraphics[scale=0.1]{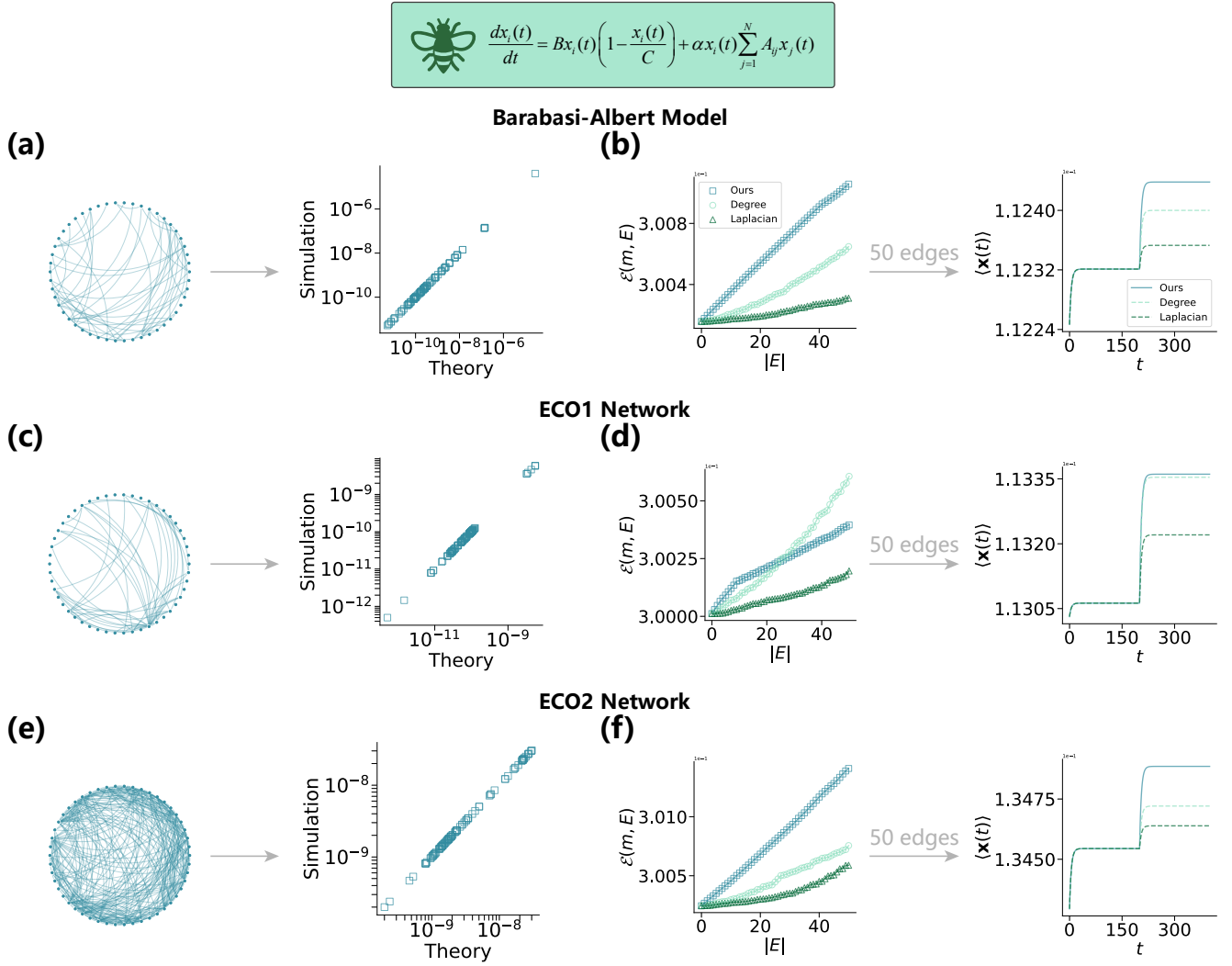}
\caption{\label{figure:6}\textbf{Simulations on mutualistic dynamics.} \textbf{(a)}. Simulation on Barab\'asi-Albert model and mutualistic dynamics proposed in Eq. (\ref{equ:M:1}) with $B=0.2,C=0.1,\alpha=0.01$. Our theoretical centrality $\mathcal{S}(m,E)$ v.s. change in Energy $\Delta \mathcal{E}(m,E)$ for fixed perturbed node on mutualistic dynamics \textbf{(b)} Effect of inhibiting edges ranked by different centrality on centrality of motifs $\mathcal{E}(m,E)$ and average state $\langle\mathbf{x}(t)\rangle$. \textbf{(c-d)} ECO1 network. \textbf{(e-f)} ECO2 network.}
\end{figure*}

\section{Applications to Mutualistic Dynamics}
\par Research in ecosystems is a popular topic in complex networks. With the development of human society and the widespread acceptance of ecological values, maintaining the stability of ecosystems and protecting biodiversity have become issues of concern for governments and relevant departments. Empirically, it is believed that the more complex the system species, the more robust the system is. However, Robert May published a groundbreaking article\cite{May1976SimpleMM}, providing a theorem that the stability of a system decreases with the increase in network size and interaction strength. Detailed mathematical proofs reveal extremely different physical laws, indicating that the study of ecosystems requires the introduction of a large number of tools from complex networks.

\par In mathematical models, we generally consider the following three types of interactions between species in biology. The first type is predation: for example, the relationship between predator $A$ and prey $B$, where $A$ has an inhibitory effect on $B$ (denoted by $-$), and $B$ has a mutualistic effect on $A$ (denoted by $+$)\cite{Wang2024NetworkedDS}. The second type of relationship is intra-population competition: for example, the relationship between species $A$ and species $B$ due to competition for the same resources, where there is an inhibitory effect between $A$ and $B$ (denoted by $-$)\cite{Wang2024NetworkedDS}. The third type of relationship is mutualism within populations: for example, the parasitic relationship between species $A$ and species $B$, where there is a mutualistic effect between $A$ and $B$ (denoted by $+$)\cite{Wang2024NetworkedDS}. The combination and interactions of these three types of relationships form the basic impact patterns of an ecosystem.

\par In this section, we consider the mutualistic relationships between species, which are related to building stable, robust ecosystems. For example, the restoration of desert ecosystems and the protection of river/sea ecosystems both require external interventions to ensure the maintenance of mutual promotion among species. The general expression for mutualistic dynamics is\cite{May1976SimpleMM,Holling1959SomeCO}:

\begin{equation}
    \begin{aligned}
        \label{equ:M:1}
        \frac{dx_i(t)}{dt}=Bx_i(t)\left(1-\frac{x_i(t)}{C}\right)+\alpha x_i(t)\sum_{j=1}^NA_{ij}x_j(t),
    \end{aligned}
\end{equation}
in which $x_i(t)$ is the group size of specie $i$ at time $i$, $B$ is the reproducing process coefficient, $C$ is the opposite effect coefficient representing the competition due to limited resources, and $\alpha$ captures the interaction strength between two species. In artificially maintained networks, interactions between weaker species typically result in smaller values of $\alpha$, requiring strong powers to maintain the stability of the ecosystem.

\subsection{Verification of Theoretical Framework}
\label{section:M2}
\par In real ecosystems, we generally select two main approaches: introducing species and enhancing interactions between species. In the context of our framework, increasing the number of species is equivalent to adding state of nodes to the system, while strengthening interactions between species is akin to increasing the weights of edges (weakening interactions is equivalent to reducing weights or deleting edges). Therefore, edges with higher scores represent species relationships that require particular attention.

\par Simulations are preformed on the Barab\'asi-Albert model and two real networks: in which these two networks are collected from the symbiotic interactions in Carlinville Illinois\cite{Marlin2001TheNB}. The interactions matrix $M$ between plants and pollinators is stored by a matrix with size $456\times 1429$, which can be expressed as a bipartite graph containing 456 plants and their 1429 pollinators. The first network is obtained from $M$ containing 456 nodes, and its academic name is the plant-plant mutualistic network(ECO1), there exists an edge between two plant $i$ and $j$ if density $\sum\limits_{k=1}^{456}\frac{M_{ki}M_{kj}}{\sum\limits_{\ell=1}^NM_{k\ell}}$ is larger than a threshold $\eta=0.3$. The second is the pollinator-pollinator network(ECO2) containing 1429 nodes, edges are determined with density $\sum\limits_{k=1}^{1429}\frac{M_{ik}M_{jk}}{\sum\limits_{\ell=1}^NM_{\ell k}}$. Obviously, ECO1 and ECO2 are all symmetric networks. We included the weights of edges in the suggested ecosystem and simulated the observed change in energy $\Delta \mathcal{E}(m,E)$. The simulation results have been illustrated in Figure \ref{figure:6} and clearly show that our theoretical results can perfectly predict $\Delta \mathcal{E}(m,E)$ in ecosystems and validates the correctness of our theoretical framework.

\subsection{Advice to Ecological Protection}
\par In some real-life ecosystems, simply increasing the number of organisms may not necessarily yield optimal results and can be cost-ineffective, as seen in the case of the failure of Biosphere 2. Therefore, it is necessary to enhance interactions between species during the process of species introduction to strengthen system stability. Methods to enhance interactions include constructing biological exchange pathways and protecting habitats. Efficient ecological conservation measures can protect fragile ecosystems to the maximum extent at minimal cost, safeguarding the living environment for human.

\par Then, we conducted simulations on the theoretical network and two real networks mentioned in Section \ref{section:M2}. We calculated $\mathcal{E}(m,E)$ over the size of $E$ and $\langle \mathbf{x}(t)\rangle$ after adding the weight of certain edges over time $t$, where $\mathcal{E}(m,E)$ can signify the efficiency of increasing species diversity. These results were compared with both degree centrality and Laplacian centrality and are depicted in Figure \ref{figure:6}. It is evident that in almost all cases our results consistently outperformed the other two centralities. The significant improvement in control under a small number of edges indicates that our centrality offers a distinct advantage in safeguarding fragile ecosystems.
\section{Applications to Regulatory Dynamics}
\par Gene regulation is a very important biochemical process, which is the mechanism that controls gene expression within an organism. It mainly involves the transcription of genes and the translation of mRNA. For multicellular organisms, gene regulation is closely related to cell differentiation and individual development. In mathematical models, we generally use the Michaelis-Menten dynamics to characterize the basic process of gene regulation. The basic expression is\cite{Alon2019AnIT,Karlebach2008ModellingAA,Barzel2011BinomialME}:
\begin{equation}
    \begin{aligned}
        \label{equ:R:1}
        \frac{dx_i(t)}{dt}=-Bx_i^a(t)+\alpha\sum_{j=1}^NA_{ij}\frac{x_j^b(t)}{1+x_j^b(t)},
    \end{aligned}
\end{equation}
in which $x_i(t)$ the level of gene expression of gene $i$ at time $t$. The different values of the parameter $a$ in the self-dynamics $F(x) = -Bx^a$ determine various biochemical processes. For instance, $a = 1$ corresponds to the degradation process, while $a = 2$ corresponds to the dimerization process. The interaction dynamics $H_2(x) = \frac{x^b}{1+x^b}$ with $b>0$ is known as the Hill function, which records the cooperation level of gene $i$ by gene $j$\cite{Karlebach2008ModellingAA}. It possesses properties such as $H_2(x\to 0) = 0$ and $H_2(x\to\infty) = 1$, making it a switch-like dynamic. This function is similar to a gate function in neural networks. Due to the specific properties of the Hill function, only a subset of genes in gene interactions will actually have an effect\cite{PhysRevLett.130.097401}.

\begin{figure*}[tbp]
\centering
\includegraphics[scale=0.1]{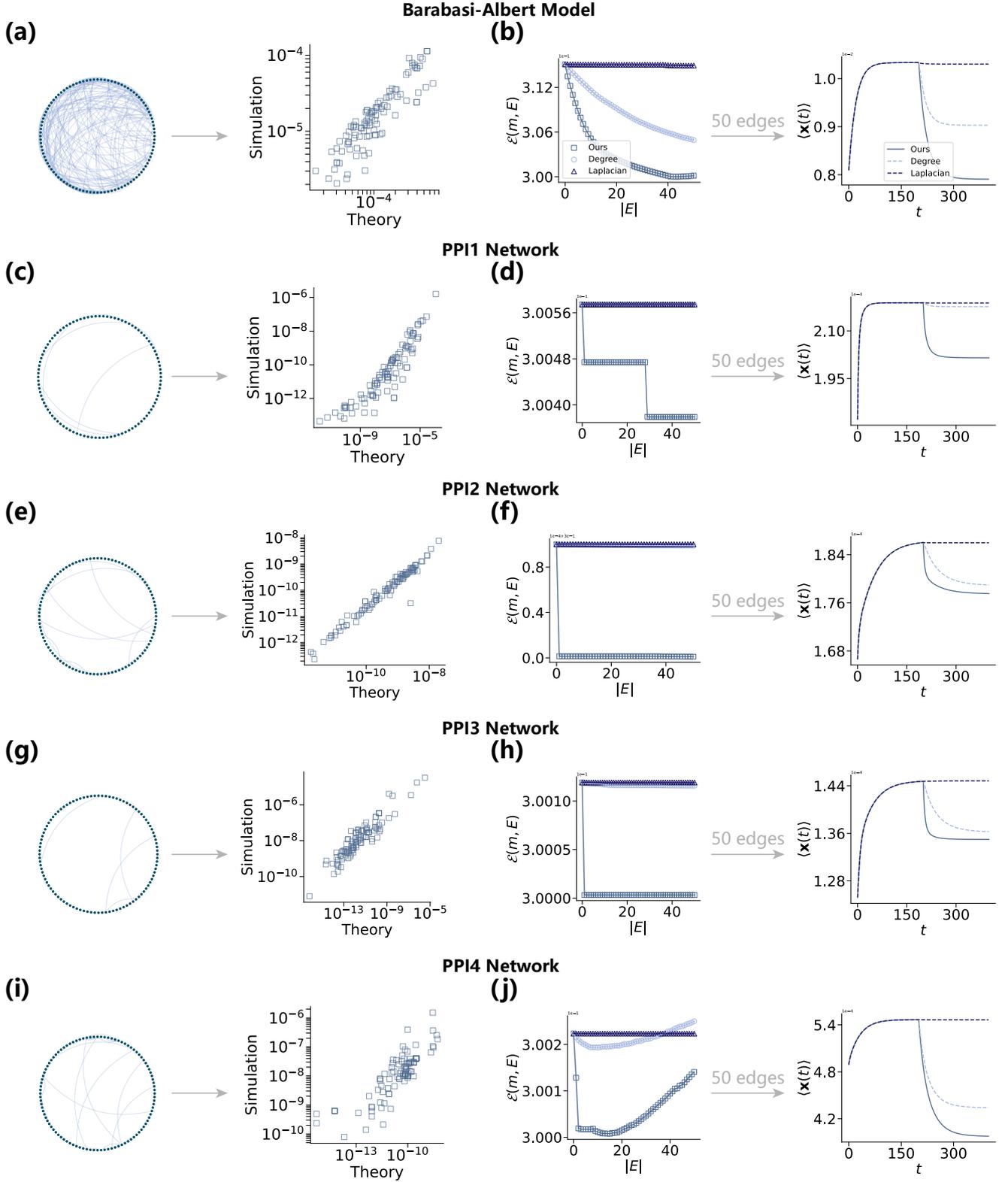}
\caption{\label{figure:7}\textbf{Simulations on regulatory dynamics.} \textbf{(a)}. Simulation on Barab\'asi-Albert model and regulatory dynamics proposed in Eq. (\ref{equ:R:1}) with $B=0.5,a=1.1,b=1,\alpha=0.01$. Our theoretical centrality $\mathcal{S}(m,E)$ v.s. change in Energy $\Delta \mathcal{E}(m,E)$ for fixed perturbed node on mutualistic dynamics \textbf{(b)} Effect of inhibiting edges ranked by different centrality on centrality of motifs $\mathcal{E}(m,E)$ and average state $\langle\mathbf{x}(t)\rangle$. \textbf{(c-d)} PPI1 network. \textbf{(e-f)} PPI2 network. \textbf{(g-h)} PPI3 network. \textbf{(i-j)} PPI4 network.}
\end{figure*}

\subsection{Verification of Theoretical Framework}
\label{section:R2}
\par We still integrate the mechanisms for gene regulation into our framework through the following mappings. The control of gene expression levels can be regard as the change of node states. Controlling the concentration of a certain protein in the environment can regulate the degree of interaction between two genes, i.e., altering the weights of edges. Through these two methods, we can achieve control over the gene regulation process.
\par Simulations are preformed on the Barab\'asi-Albert model and four real networks: the first is PPI1(academic name as the protein-protein interaction network of yeast), consisting of 1647 nodes and 5036 edges\cite{Yu2008HighQualityBP}. The second is PPI2(academic name as the human protein-protein interaction network), consisting of 2035 nodes and 13806 edges\cite{Rual2005TowardsAP}. The third is PPI3(academic name as the Arabidopsis thaliana protein-protein interaction network), consisting of 2938 nodes and 7720 edges\cite{Ikehara2017CharacterizingTS}. The fourth is PPI4(academic name as the Rattus norvegicus gene-protein interaction network), consisting of 2350 nodes and 3484 edges\cite{Domenico2015StructuralRO}. We removed certain edges in the given protein-protein or gene-protein network and simulated the resulting change in energy $\Delta \mathcal{E}(m,E)$. The simulation outcomes are depicted in Figure \ref{figure:7}, demonstrating that our theoretical predictions accurately match $\Delta \mathcal{E}(m,E)$ in the gene network, thereby confirming the validity of our theoretical framework.

\subsection{Advice to Gene Regulation}
\par In this section, we continue to utilize edge manipulation to regulate gene expression. We removed specific edges to evaluate the gene expression level and monitor alterations in expression levels. Aligned with our framework, we conducted simulations on the theoretical network and four real networks referenced in Section \ref{section:R2}. We computed $\mathcal{E}(m,E)$ across the edge set $E$ and $\langle \mathbf{x}(t)\rangle$ by incorporating the weights of select edges over time $t$, where $\mathcal{E}(m,E)$ can indicate variations in gene expression levels. These findings were compared with both degree centrality and Laplacian centrality and are illustrated in Figure \ref{figure:7}. It is obvious that our results consistently surpassed the other two centralities in nearly all instances. The enhancement in control with a limited number of edges suggests that our centrality provides a unique advantage in gene regulation.
}
\section{Discussion and Outlook}
\par In this paper, we propose a centrality measure for identifying key motifs in dynamical complex systems. This measure is based on the definition of perturbation energy, which is determined by the elements in the Jacobian matrix. It considers both the topological connections and dynamic properties of the system, revealing the propagation and stability characteristics of the system. Moreover, our centrality only requires the calculation of matrix elements, significantly improving the precision of predicting key motifs without imposing a significant computational burden. To validate our theoretical framework, we conducted simulations on real dynamic systems and integrated them with real world scenarios. Our findings validate the effectiveness and efficiency of our centrality measure for dynamic control, providing significant implications for the formulation of epidemic prevention, ecological protection, gene regulation control, and policies for managing public opinion.

\par However, our framework is effective for small perturbations and fixed interaction matrices, large perturbations can destabilize the system, making it challenging to accurately estimate system states. Addressing large perturbations necessitates the use of nonlinear dynamics, which requires advanced mathematical tools\cite{Bianconi2023ComplexSI}. Additionally, for time-varying systems where the interaction matrix depends on time $t$, it may be possible to use the Jacobian matrix at a specific moment to replace the matrix in our proposed centrality. However, the implications of this substitution require further exploration and a deeper understanding of time-varying systems.

\section*{Declaration of Competing Interest}
The authors declare no competing interests.

\section*{Acknowledgments}
\par {We would like to thank the referees for their constructive comments and suggestions.} This work is partly supported by the National Natural Science Foundation of China (Nos.12371354, 12161141003, 11971311) and Science and Technology Commission of Shanghai Municipality, China (No.22JC1403600), National Key R\&D Program of China under Grant No. 2022YFA1006400 and the Fundamental Research Funds for the Central Universities, China.

\section*{Data Availability}
\par The data and code that support the findings of this study are openly available in GitHub, \url{https://github.com/QitongHu2000/Centrality-Signal-Propagation-data}.

\section*{Code Availability}
\par The data and code that support the findings of this study are openly available in GitHub, \url{https://github.com/QitongHu2000/Centrality-Signal-Propagation-main}.

\appendix
{
\section{Calculations for Mathematical Equations}
\subsection{Detailed Proof for Eq.(\ref{equ:dynamic:3.5})}
\par According to the estimation of $\Delta\mathbf{x}(\infty,m)$ proposed in Eq. (\ref{equ:dynamic:3}), we use the Sherman-Morrison formula $(A+uv^T)^{-1}=A^{-1}-\frac{A^{-1}uv^TA^{-1}}{1+v^TA^{-1}u}$ to calculate $\left(sI-\mathbf{J}+e_me_m^T\mathbf{J}\right)^{-1}$ and consider the limit as $s$ approaches 0. In this case, there should be
\begin{equation}
    \begin{aligned}
        \label{equ:appendix:A1}
        &\lim_{s\to \infty}s\left(sI-\mathbf{J}+e_me_m^T\mathbf{J}\right)^{-1}e_m\\
        =&\lim_{s\to \infty}s(sI-\mathbf{J})^{-1}\left(I-\frac{e_me_m^T\mathbf{J}(sI-\mathbf{J})^{-1}}{1+(\mathbf{J}(sI-\mathbf{J})^{-1})_{mm}}\right)e_m\\
        =&\lim_{s\to \infty}\frac{s(sI-\mathbf{J})^{-1}e_m}{1+(\mathbf{J}(sI-\mathbf{J})^{-1})_{mm}},
    \end{aligned}
\end{equation}
in which $A_{mm}$ represents the value of matrix $A$ on the $m$-th row and $m$-th column, which is equivalent to $e_m^TAe_m$. Next, we analyze the remaining term in Eq. (\ref{equ:dynamic:3}) and consider the fact that $\lim\limits_{s\to 0}G(s)=\int_0^{\infty}f(t)dt$. This enables us to derive an accurate estimation for $\lim\limits_{s\to \infty} e_m^T\left[\Delta \mathbf{x}(0)+G(s)\right]e_m$:
\begin{equation}
    \begin{aligned}
        \label{equ:appendix:A2}
        \lim_{s\to \infty} e_m^T\left[\Delta \mathbf{x}(0)+G(s)\right]e_m&=\Delta x_m(0)+\int_{0}^\infty f(t)dt\\
        &=\Delta x_m(\infty).
    \end{aligned}
\end{equation}
By combining Eq. (\ref{equ:appendix:A1}) and Eq. (\ref{equ:appendix:A2}), and based on the definitions of $\Delta \mathbf{x}(0)$ and $G(s)$, we can derive $\left[\Delta \mathbf{x}(0)+G(s)\right]e_m=e_me_m^T\left[\Delta \mathbf{x}(0)+G(s)\right]e_m$. This allows us to obtain the precise expression for $\Delta \mathbf{x}(\infty,m)$:
\begin{equation}
    \begin{aligned}
        \Delta \mathbf{x}(\infty,m)
        &=\lim_{s\to 0}s\left(sI-\mathbf{J}+e_me_m^T\mathbf{J}\right)^{-1}e_me_m^T\left[\Delta \mathbf{x}(0)+G(s)\right]e_m\\
        &=\lim_{s\to 0}\frac{s(sI-\mathbf{J})^{-1}e_m\Delta x_m(\infty)}{1+(\mathbf{J}(sI-\mathbf{J})^{-1})_{mm}}.
    \end{aligned}
\end{equation}
Finally, taking into account the property $((sI-\mathbf{J})(sI-\mathbf{J})^{-1})_{mm}=1$, Eq. (\ref{equ:dynamic:3}) can be simplified to:
\begin{equation}
    \begin{aligned}
        \Delta \mathbf{x}(\infty,m)
        &=\lim_{s\to 0}\frac{s(sI-\mathbf{J})^{-1}e_m\Delta x_m(\infty)}{((sI-\mathbf{J})(sI-\mathbf{J})^{-1})_{mm}+(\mathbf{J}(sI-\mathbf{J})^{-1})_{mm}}\\
        &=\lim_{s\to 0}\frac{(sI-\mathbf{J})^{-1}e_m\Delta x_m(\infty)}{((sI-\mathbf{J})^{-1})_{mm}}\\
        &=\frac{\mathbf{J}^{-1}e_m}{(\mathbf{J}^{-1})_{mm}}\Delta x_m(\infty).
    \end{aligned}
\end{equation}
\subsection{Detailed Proof for Eq.(\ref{equ:energy:1})}
\par In the framework outlined in the main text, we introduce a small perturbation value $\varepsilon$ to a set of edges $E$. This perturbation impacts the Jacobian matrix $\hat{\mathbf{J}}(E)=\mathbf{J}-\Delta\mathbf{J}(E)$, where $\Delta \mathbf{J}(E)=\varepsilon \sum\limits_{(i,j)\in E}\mathbf{J}_{ij}e_ie_j^T$. We have defined the following terms: $\mathcal{E}(m)=\Vert\Delta\mathbf{x}(\infty,m)\Vert_2^2=\frac{(\mathbf{J}^{-T}\mathbf{J}^{-1})_{mm}}{(\mathbf{J}^{-1})_{mm}^2}\Delta x_m^2(\infty)$, $\mathcal{E}(m,E)=\Vert\Delta \hat{\mathbf{x}}(\infty,m,E)\Vert_2^2$, and $\Delta \mathcal{E}(m,E)=\mathcal{E}(m)-\mathcal{E}(m,E)$. Subsequently, we have:
\begin{equation}
    \begin{aligned}
        \Delta \mathcal{E}(m,E)
        =&\left(\frac{(\mathbf{J}^{-T}\mathbf{J}^{-1})_{mm}}{(\mathbf{J}^{-1})_{mm}^2}-\frac{(\hat{\mathbf{J}}(E)^{-T}\hat{\mathbf{J}}(E)^{-1})_{mm}}{(\hat{\mathbf{J}}(E)^{-1})_{mm}^2}\right)\Delta x_m^2(\infty),
    \end{aligned}
\end{equation}
as proposed in the main text, we define $\Delta G_1(E)=\mathbf{J}^{-T}\mathbf{J}^{-1}-\hat{\mathbf{J}}^{-T}(E)\hat{\mathbf{J}}^{-1}(E)$ and $\Delta G_2=\mathbf{J}^{-1}-\hat{\mathbf{J}}^{-1}(E)$. And we can have:
\begin{equation}
    \begin{aligned}
        &\Delta \mathcal{E}(m,E)\\
        =&\left(\frac{(\mathbf{J}^{-T}\mathbf{J}^{-1})_{mm}}{(\mathbf{J}^{-1})_{mm}^2}-\frac{(\hat{\mathbf{J}}(E)^{-T}\hat{\mathbf{J}}(E)^{-1})_{mm}}{(\mathbf{J}^{-1})_{mm}^2}\right)\Delta x_m^2(\infty)\\
        &\quad+\left(\frac{(\hat{\mathbf{J}}(E)^{-T}\hat{\mathbf{J}}(E)^{-1})_{mm}}{(\mathbf{J}^{-1})_{mm}^2}-\frac{(\hat{\mathbf{J}}(E)^{-T}\hat{\mathbf{J}}(E)^{-1})_{mm}}{(\hat{\mathbf{J}}(E)^{-1})_{mm}^2}\right)\Delta x_m^2(\infty)\\
        =&\left(\frac{(\Delta G_1(E))_{mm}}{(\mathbf{J}^{-1})_{mm}^2}-2\frac{(\hat{\mathbf{J}}(E)^{-T}\hat{\mathbf{J}}(E)^{-1})_{mm}(\Delta G_2(E))_{mm}}{(\mathbf{J}^{-1})^3_{mm}}\right)\Delta x_m^2(\infty),
    \end{aligned}
\end{equation}
in which for the second term, we apply the Lagrange mean value theorem $\frac{1}{a^2}-\frac{1}{b^2}=-\frac{2}{\theta^3}$ with $\theta\in [a,b]$, where $a=(\mathbf{J}^{-1})_{mm}^2$ and $b=(\hat{\mathbf{J}}(E)^{-1})_{mm}^2$. Therefore, we can approximate $\theta\approx a=(\mathbf{J}^{-1})_{mm}^2$ and $(\hat{\mathbf{J}}(E)^{-T}\hat{\mathbf{J}}(E)^{-1})_{mm}\approx (\mathbf{J}^{-T}\mathbf{J}^{-1})_{mm}$ when $\varepsilon$ is very small. Consequently, we can derive the simplified expression for $\Delta \mathcal{E}(m,E)$:
\begin{equation}
    \begin{aligned}
        &\Delta \mathcal{E}(m,E)\\
        =&\frac{(\mathbf{J}^{-T}\mathbf{J}^{-1})_{mm}}{(\mathbf{J}^{-1})_{mm}^2}\left(\frac{(\Delta G_1(E))_{mm}}{(\mathbf{J}^{-T}\mathbf{J}^{-1})_{mm}}-2\frac{(\Delta G_2(E))_{mm}}{(\mathbf{J}^{-1})_{mm}}\right)\Delta x_m^2(\infty).
    \end{aligned}
\end{equation}
}
\bibliographystyle{elsarticle-num}

\end{document}